\documentclass[12pt, preprint]{aastex}

\shorttitle{IR Spectrum of 2003 EL61's Satellite}
\shortauthors{Barkume et al.}

\begin{document}
\title{Water Ice on the Satellite of Kuiper Belt Object 2003 EL61}
\author{K.M Barkume,  M.E. Brown, and E.L. Schaller}
\affil{Division of Geological and Planetary Sciences, California Institute
of Technology, Pasadena, CA 91125}

\email{barkume@caltech.edu, mbrown@caltech.edu, emily@caltech.edu}

\begin{abstract}

We have obtained a near infrared spectrum of the brightest satellite of the large Kuiper Belt Object, 2003 EL61.  The spectrum has absorption features at 1.5 and 2.0 microns, indicating that water ice is present on the surface. We find that the satellite's absorption lines are much deeper than water ice features typically found on Kuiper Belt Objects.  We argue that the unusual spectrum indicates that the  satellite was likely formed by impact and not by capture.

\end{abstract}

\keywords{comets: general -- infrared: solar system -- minor planets}

\section{Introduction}
Satellites in the Kuiper Belt provide unique insights into the internal properties of icy bodies as well as the early dynamical history of the trans-Neptunian region. To date, over a dozen satellites have been found around Kuiper Belt Objects (KBOs) and the overall satellite fraction is estimated to be 11$\pm ^5_2 $\% (Stephens et al., 2005). The first KBO satellites discovered (excluding the Pluto-Charon system) were found to have high eccentricities, large orbital separations, and have similar brightness to the primary, indicating that the satellites are nearly as large as their parent.  Various capture mechanisms were proposed to account for the high specific angular momentum of these systems (Goldreich et al., 2002). Recently,  Brown et al. (2006) have found several satellites around the largest known KBOs, and these satellites appear to be different than those discussed above. The satellite fraction appears to be higher around large KBOs and, like the Pluto-Charon system, the satellites have lower fractional brightness compared to the primary. The orbit of the brightest satellite around 2003 EL61 is nearly circular and  Brown et al. (2006) suggest that these systems likely formed by a different mechanism and may be the result of giant impacts (McKinnon 1989, Canup 2005). The density, albedo, and surface composition of these satellite systems may help to clarify the origins of satellites in the Kuiper belt.

Near infrared spectroscopy is a particularly well suited method for remotely studying the surface composition of KBOs. Many of the abundant species in the outer solar system, including CH$_4$, CO$_2$, and H$_2$O ices, have absorption features in the near IR. A small amount of water ice has been observed on a number of KBOs  and some show evidence for the presence of other ices (Fronasier et al. 2004 and references therein).  The spectra of satellites may shed light their origin. Captured objects are likely to have spectra similar to other KBOs, but satellites formed from other mechanisms could show unusual or altered surface compositions. 

2003 EL61 is currently the third brightest known KBO and is elliptical in shape with axes of 1960  and 2500 km (Rabinowitz et al., 2006). Its satellites offer an excellent opportunity to study the formation and dynamics of KBO binaries. The brighter satellite, S/2005~(2003~EL61)~1, has a period 49.12 $\pm$ 0.03 days and a maximum separation of 1.4 arcseconds (Brown et al. 2005). The orbit is nearly circular, with an eccentricity of 0.05.  The inner satellite, S/2005~(2003~EL61)~2, appears to have an orbit of 34.7 $\pm$ 0.1 days, though there is currently insufficient data to reliably fit a non-circular orbit.  The inner satellite has a fractional brightness of 1.5\% compared to 2003 EL61, making it difficult to study with spectroscopic techniques. The outer, brighter satellite has a fractional brightness of 5\% and is a viable target for low resolution infrared spectroscopy. We report here on the near-IR (J through K band) spectrum of  2003 EL61's brighter satellite. We compare the spectrum of this satellite to 2003~EL61's as well as to other KBOs.  We discuss possible formation scenarios for the satellite and find that the surface composition is consistent with formation by giant impact.

\section{Observations}

Observations of 2003 EL61 were obtained on the Keck Telescopes on UT 2005 April 26 and 27 using NIRC, the facility infrared camera (Matthews \& Soifer 1994).  J band images revealed that the brighter satellite was near its maximum separation on these nights, with an average distance of 1.4 arcseconds from the primary. The average seeing was 0.5 arcseconds on both nights. Signal from the satellite was easily separated from the primary in the raw imaging and spectral data. Figure 1 shows the J band image of 2003 EL61 and the outer satellite. The dimmer satellite was not detected.

Spectra of 2003 EL61 and the brighter satellite were taken simultaneously by placing them in a .52 $\times$ 38 arcsecond slit. We obtained a spectrum in first order from 1.0 to 1.5 $\mu$m by placing a 150 lines mm$^{-1}$ grism and a J through H sorting filter in the light path.  A  1.4-2.5  $\mu$m  spectrum was obtained using a 150 lines mm$^{-1}$ grism and an H-through-K band sorting filter. Our spectral resolution in J-H is approximately $\lambda/\Delta \lambda \approx$ 132 and in H-K the resolution is  $\approx$ 162. The spectra were acquired by taking 200 second exposures at five set dithered locations with a 5 arcsecond separation between offset positions. The telescope guided at the predicted rate of motion for 2003 EL61. The total integration time for each night was 4000 seconds in H-K and 1000 seconds in J-H.  Nearby calibration stars were also observed each night at a range of air masses similar to those of 2003 EL61. 

\section{Data Reduction}
The satellite's spectrum was marginally detected in the 200-second exposures. We hand selected data in which the satellite spectum was most clearly separated from the primary in the raw data for further analysis. In the H-K band, a total of 8 spectra from the first night and 10 from the second night were selected, reducing the  total integration time to 3600 seconds. In the J band, 3 spectra from first night and 4 from the second were used, for a total integration 1400 seconds.  Data reduction was performed on the individual spectra similarly to the procedures described in Brown (2000).  After dividing each spectrum by a flat field, a majority of the sky background was removed by differencing adjacent pairs of spectra. The residual sky background was measured in 20-pixel wide swaths above and below both 2003 EL61 and the satellite. The average of the residual sky backgrounds was subtracted from every row of the spectrum. The spectra were then shifted row-wise to place 2003 EL61 in the center of the array and the individual spectra at similar air mass were added together to increase the signal from the satellite. 

2003 EL61 contributes a few percent of the flux detected at the location of satellite. To remove this contamination, a median profile of 2003 EL61 was created in the spatial direction for each night. The profile of the data was then reflected across the point of maximum flux. At each column, the profile was then scaled by the flux from 2003 EL61 in that column and subtracted from the coadded data. The spectrum of the satellite was extracted by summing over 5 rows centered at the location of the satellite.   The final spectra from were then divided by a calibration star spectrum to remove the solar spectrum and the resulting reflectance spectra were averaged.  The spectra were smoothed to increase signal-to-noise.  The spectral data was averaged at every fourth pixel using a Gaussian weighting function with a full width half maximum of 8 pixels in both the J-H and H-K band data, resulting in an oversampled spectrum with a resolution of  $\lambda/\Delta \lambda \approx$ 33 in the J-H band data and 41 in the H-K band data.

For comparison, the spectrum of 2003 EL61 was extracted in an identical manner to the outer satellite.  The H-K band data for both objects was normalized by the flux at 1.7 $\mu$m. The J band overlaps with the H-K band data and a scaling factor for the J-H data was determined by fitting the 2003 EL61's J-H and H-K band data, which has higher signal-to-noise. The scaling factor was then used to scale the J-H  spectrum for the satellite.

\section{Discussion}
The satellite spectrum shows strong absorption features at 1.5 and 2 microns, which are consistent with absorption due to water ice (Fig \ref{fig2}). However, the resolution is too poor to detect the 1.65 $\mu$m crystalline water ice absorption feature seen on 2003 EL61 (Trujillo et al., submitted). The depth of the water lines suggests that much of the satellite's surface is covered with smooth water ice.  
 We calculated a model water ice spectrum using laboratory optical constants from Grundy \& Schmitt (1998), assuming a temperature of 50 K. Our model uses Hapke theory (Hapke 1981) to transform the optical constants to a reflectance spectrum. We vary the grain size of ice in our models and find that the grain size of 250 microns provides a good fit to the absorption lines at 1.5 and 2 microns.  A better fit may be achieved with the addition of a second blue component or a multiple grain size water ice model.

The absorption features in the spectrum of  the satellite are significantly deeper than water ice features typically observed on KBOs. The ratio of flux at 1.7 $\mu$m to the flux at 2.0 $\mu$m is $\sim$ 80\%. Most KBOs with water ice features have depths of less than $\sim$20\% and the deepest absorptions ($\sim$60 \%) are found on 2003~EL61, Charon and 1996~TO66 (Brown et al. 2000, Fronasier et al 2004, Dotto et al 2003 and references therein).  As such, the brighter satellite of 2003 EL61 appears to be an unusual object in the Kuiper belt. Captured objects are likely to resemble the parent population where they are formed, thus the satellite's spectrum suggests that formation through capture is unlikely.

An alternative to capture is that the satellite could have formed via impact.  Models of the Charon-forming impact by Canup (2005) are consistent with many of the features of the 2003 EL61 system.  In these models, Charon is formed from a grazing impact where the impactor remains largely intact and creates a large satellite.  However, the models also show that a more direct impact creates a disk around the target body and smaller satellites can eventually coalesce from the disk material.  The small satellites around 2003 EL61 are more likely to be formed by the latter kind of impact than by the Charon-type grazing impacts. In models where a disk was created, the satellite preferentially forms from the less dense disk material. In the outer solar system, this trend will lead to water-enriched satellites around KBOs,  which may account for the strong water ice features seen in the satellite's spectrum.   Direct impacts can also spin up the target body, which is consistent with the high rotation rate observed for 2003 EL61 (Rabinowitz et al., 2005).   2003 EL61 has a mean density of 2.6-3.4 g/cm$^3$, indicating  that it has lost most of its volatiles (Rabinowitz et al. 2005). Further modeling is needed to show if a high energy impact can eject a substantial amount of ice off of 2003 EL61 to account for its high rock-to-ice ratio.  In sum, 2003 EL61 and its brighter satellite exhibit many of the features predicted by Kuiper belt impact models. Further analysis of other satellites may provide a greater understanding of the dynamical history of the early Kuiper belt.

{\it Acknowledgments:}
 
This research is funded by the California Institute of Technology and is also supported by the NASA Planetary Astronomy program. Data presented herein were obtained at the W. M. Keck Observatory, which is operated as a scientific partnership  among the California Institute of Technology, the University of California, and the National Aeronautics and Space Administration. The observatory was made possible by the generous financial support of the W. M. Keck Foundation.  The authors thank Chad Trujillo, David Rabinowitz, and Robin Canup for their helpful discussions regarding this paper. We also thank Antonin Bouchez for his help during observations.

\eject

\begin{figure}
\centering
	    \includegraphics[height=5in]{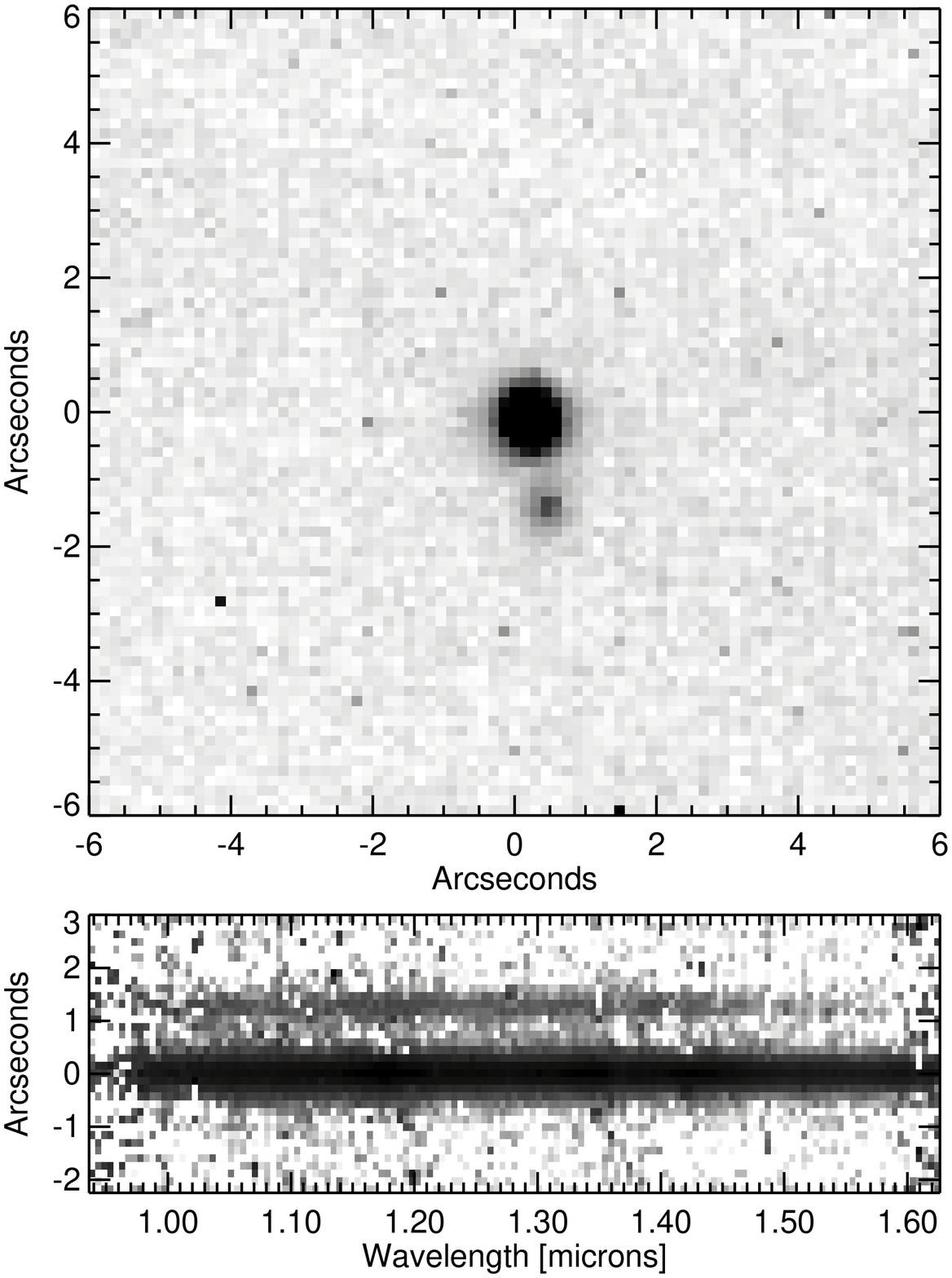}
	    \caption{The J-band band image and raw spectral data of 2003 EL61 and its outer satellite from UT 2005 April 27.  The satellite was near its maximum separation of 1.4 arcseconds. The spectral data show that the satellite's spectrum is separated from 2003 EL61. \label{fig1}}
	\end{figure}

\clearpage

\begin{figure}
\centering
	    \includegraphics[height=5in]{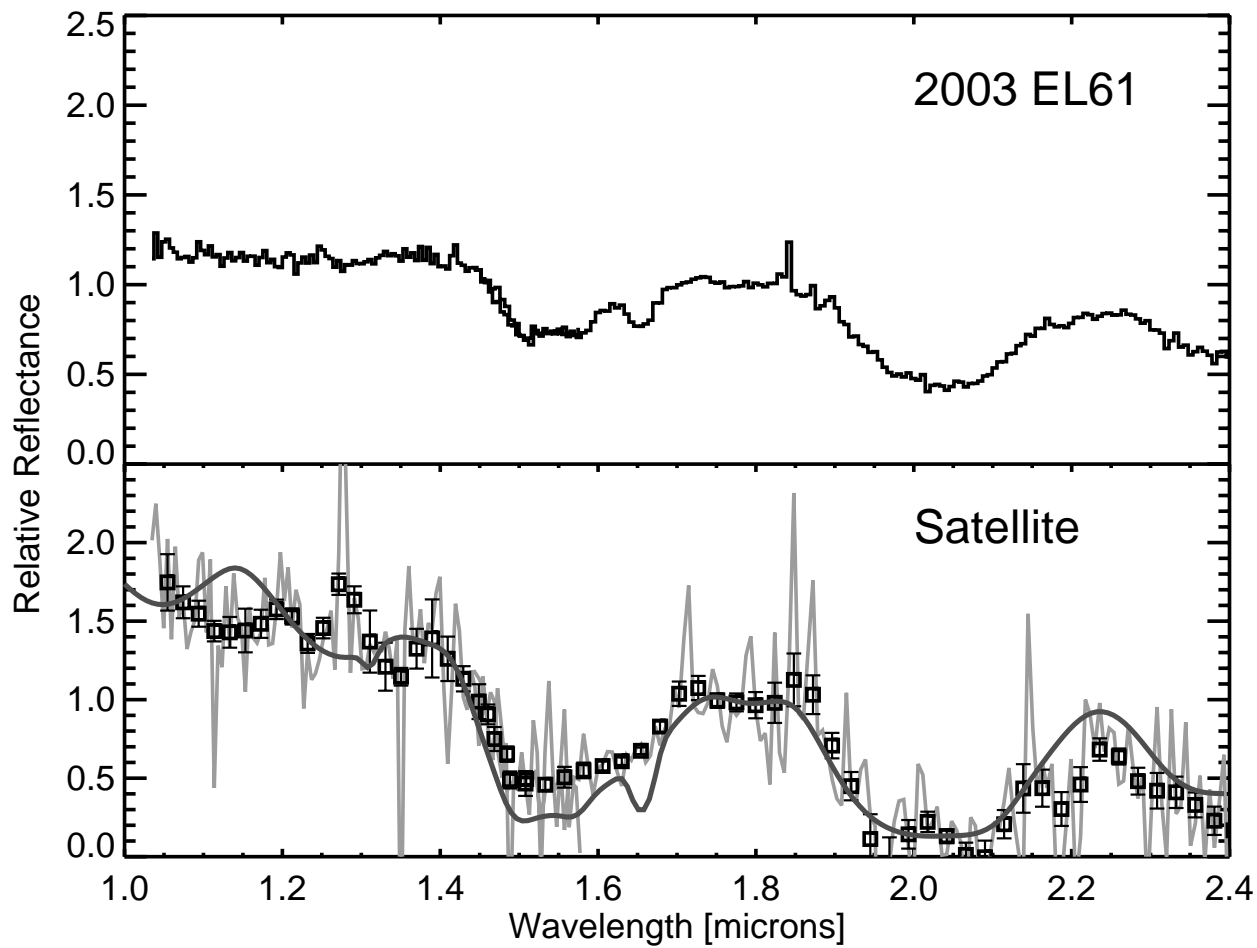}

	    \caption{The 1-2.4 $\mu$m spectrum of 2003 EL61 (top) and its outer satellite (bottom). Both spectra are scaled to the flux at 1.7 $\mu$m. The light grey line is the full resolution spectrum of the satellite and the boxes are the Gaussian smoothed data with error bars. The dark grey line shows our best fit water ice model with a grain size of 250 $\mu$m.  Unlike 2003 EL61, there is no evidence for the crystalline water ice line at 1.65 $\mu$m in the satellite's spectrum. A slight blue component improves the model fit, but there is no evidence for other common ice species. \label{fig2}}
	\end{figure}

\end{document}